# Influence of Pt/Ru ratios on the oxidation mechanism of MCrAlYTa coatings modified with Pt-Ru overlays

Majid Hosseinzadeh, Erfan Salahinejad[*]

*Faculty of Materials Science and Engineering, K.N. Toosi University of Technology, Tehran, Iran*

[*] Corresponding Author's Email Address: <salahinejad@kntu.ac.ir>

## Abstract

This study investigates the influence of varying Pt/Ru ratios on the oxidation mechanism of NiCoCrAlYTa coatings with electrodeposited, vacuum-annealed Pt-Ru overlays. Weight change measurements, scanning electron microscopy/energy dispersive X-ray spectrometry (SEM/EDS), X-ray diffraction (XRD), and X-ray photoelectron spectroscopy (XPS) were used for high-temperature oxidation analyses, showing superior resistance with higher Pt contents. This was attributed to the creation of a denser, thinner, and more homogeneous layer of alumina ($\alpha$-$Al_2O_3$) in the thermally-grown oxide (TGO) layer. On the contrary, an increase in Ru contents led to the development of other oxides and microcracks along with alumina in the TGO layer, undermining oxidation protection. The accommodation of Ti and Ta, in the minimally-deteriorative form of carbide, along with Y into the TGO layer with increasing Pt contents further enhanced oxidation resistance. In addition to the explored significant impact of the Pt/Ru ratio on oxide scale characteristics and oxidation resistance, the lower cost of Ru compared to Pt suggests the potential for designing cost-effective systems through optimized Pt/Ru ratios and microstructural engineering.

# 1. Introduction

MCrAlY (with M representing Ni and/or Co) coatings are commonly utilized within thermal-barrier coatings (TBCs) on hot-section components in aerospace and terrestrial gas turbine engines to offer high-temperature oxidation and corrosion resistance. The superior efficacy of this protective layer, used as either an overlay or a bond coat, is mainly ascribed to the development of a dense, adherent, stable, slowly growing, and continuous alumina ($\alpha$-$Al_2O_3$) layer on the surface, called thermally-grown oxide (TGO). The longevity of MCrAlY coatings is predominantly influenced by the extent of aluminum consumption within the coating, driven by surface oxidation and diffusion into the underlying substrate. Additionally, issues like cracking, rumpling, and spallation at the TGO scale culminate in TBC degradation [1, 2]. These challenges highlight the necessity of modifying traditional MCrAlY coatings to enhance the system's lifetime and reliability.

The ability of MCrAlY coatings to resist oxidation can be enhanced through several mechanisms: 1) enhancing TGO layer adhesion to the coating surface to prevent spallation, 2) reducing the TGO growth rate to delay attaining the critical thickness, and 3) minimizing the reciprocal diffusion of the alloy constituents between the coating and substrate [3]. A well-documented and effective method for modifying these coatings is the incorporation of platinum (Pt). The advantageous impacts of Pt additions on MCrAlY coatings encompass: (1) reducing growth stress within the oxide scale [4, 5], (2) enhancing aluminum affinity and facilitating its uphill diffusion [6, 7], (3) inhibiting void formation beneath the oxide scale [8, 9], (4) promoting outward aluminum diffusion while restricting the diffusion of alloying elements, thereby reducing the development of non-protective oxides and supporting the creation of a pure alumina layer [10, 11], and (5) promoting alumina formation and enhancing alumina layer adhesion [12, 13]. However, Pt-modified coatings are susceptible to rumpling in the course of thermal cycling owing to the mismatch in thermal expansion coefficients between TGO and its underlaying coating. This phenomenon results in the considerable out-of-plane displacement of TGO, which can initiate cracks within or between the multilayers of the TBC system, ultimately giving rise to the spallation of the TBC [14, 15]. Also, the high cost of Pt remains a barrier for its widespread application, driving research into partial or complete alternatives that provide comparable outcomes.

Ruthenium (Ru) incorporation is a viable option for enhancing the high-temperature performance of bond coats, where Ru offers chemical properties similar to Pt at a more cost-effective price while reducing rumpling. These enhancements are attributed to: 1) promoting the formation of transient alumina [16, 17], 2) forming thinner $\alpha$-$Al_2O_3$ scales [18, 19], 3) decreases the Al activity, which slows the growth rate of alumina [20], and 4) increasing the creep and rumpling resistance of the bond coat [21, 22]. A recent preliminary study by the same authors provided observations on the oxidation protection of pure Ru, pure Pt, and a Pt-Ru alloy (66Pt-34Ru in wt%) overlays on an MCrAlY bond coat [23], without exploring a broader range of Pt/Ru ratios or presenting a more comprehensive analysis of the involved mechanisms. The present work systematically examines a broader range of Pt/Ru ratios, particularly compositions across distinct regions of the Pt-Ru phase diagram [24], including a Pt-rich solid solution, a Ru-rich solid solution, and a biphasic composition of Pt and Ru. By providing insights into the potential of these additives, both individually and in combination, this study aims to explore possible synergistic effects and inform the design of more efficient and cost-effective bond coats with enhanced oxidation protection.

## 2. Experimental procedure

### 2.1. Sample preparation

NiCoCrAlYTa powder (Amdry 997, Sulzer Metco), with a mean particle size of approximately 30 μm, was employed as the coating material for grit-blasted CMSX-4 superalloy specimens, applied via a high-velocity oxygen-fuel (HVOF) spraying apparatus (GTV Verschleiss- Schutz GmbH, Germany). Table 1 lists the nominal composition of the CMSX-4 superalloy and NiCoCrAlYTa powder used in this study. The operational parameters for HVOF deposition are also tabulated in Table 2.

**Table 1.** Chemical composition of the CMSX-4 superalloy and the Amdry997 powder (wt%)

|         | Ni   | Co   | Cr   | Al  | Y   | Ta  | Ti  | W   | Mo  | Re  | Hf  |
|---------|------|------|------|-----|-----|-----|-----|-----|-----|-----|-----|
| CMSX-4  | 54.5 | 9.5  | 6.4  | 5.7 | 6.5 | 6.5 | 1.0 | 6.3 | 0.6 | 2.9 | 0.1 |
| Amdry997| 43.9 | 23.0 | 20.0 | 8.5 | 0.6 | 4.0 | -   | -   | -   | -   | -   |

**Table 2.** HVOF parameters used for depositing NiCoCrAlYTa on the CMSX-4 substrate

| Parameters                                        | Value |
|---------------------------------------------------|-------|
| Oxygen flow rate (standard liter per minute, SLPM)| 850   |
| Kerosene flow rate (SLPM)                         | 0.5   |
| Carrier gas (Ar) flow rate (SLPM)                 | 7     |
| Spraying distance (mm)                            | 300   |
| Powder feed rate (g/min)                          | 90    |
| Spray gun traversing velocity (mm/s)              | 3     |

The electrodeposition procedure of Pt on the NiCoCrAlYTa-coated samples was performed in an alkaline solution containing $Pt(NH_3)_2(NO_2)_2$ at 90 °C, with a platinum level of 10 g/L, pH of 11, and the current density of 1 A/dm² for 180 min. The electrodeposition process for Ru was executed within a solution containing $RuCl_3$ at 70 °C, demonstrating a substantial Ru concentration of 5 g/L and pH of 1, at a current density of 5 A/dm² for 60 min. The electrodeposition bath for the Pt/Ru alloy also comprised stoichiometrically balanced concentrations of $Pt(NH_3)_2(NO_2)_2$ and $RuCl_3$, specifically 5-10 g/L of the former and 5 g/L of the latter at the current densities of 2, 3, and 4 A/dm² for 90 min, for the nominal compositions of 70Pt-30Ru, 30Pt-70Ru and 10Pt-90Ru in wt%, respectively, at pH of 3. Then, the specimens underwent vacuum annealing at 1080 °C for 6 h. Figure 1 illustrates a detailed overview of the different coated samples and manufacturing processes used.

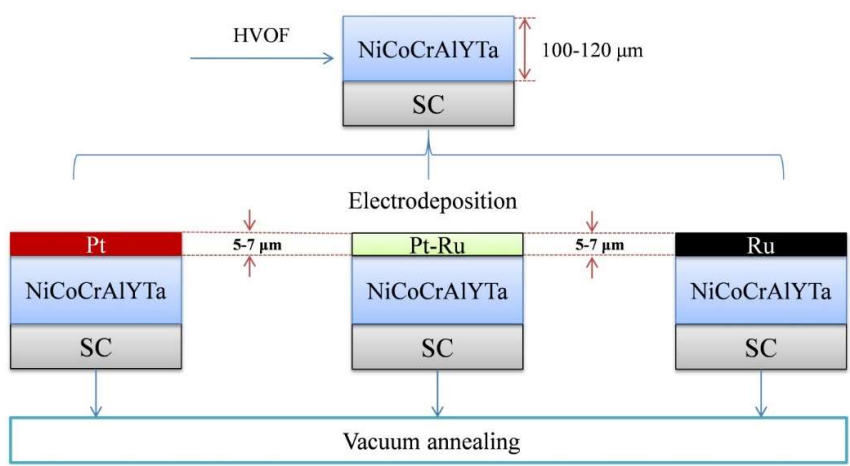

**Figure 1**. Schematic illustration of the preparation process (SC stands for single-crystal CMSX4 superalloy).

## 2.2. Structural characterization

The top and cross-sectional morphologies of the samples were systematically analyzed utilizing field-emission electron microscopy (FESEM, TESCAN, MIRA3) in both secondary electron (SE) and backscattered electron (BSE) imaging modes, which was complemented by energy dispersive X-ray spectrometry (EDS).

## 2.3. High-temperature oxidation characterization

Isothermal oxidation experiments were performed in a muffle furnace at 1100 °C for a duration of up to 200 h in air. Variations in the mass of the specimens were recorded at consistent intervals utilizing an A&D analytical balance, across three separate trials. X-ray diffraction (XRD, EXPLORER, GNR) utilizing Cu Kα radiation was also utilized to characterize the phase structures formed in the samples during the oxidation tests, while the obtained morphologies and compositions were analyzed by SEM-EDS. Additionally, X-ray photoelectron spectroscopy (XPS, Bestek) was conducted using monochromatic aluminum K-alpha excitation (1486.7 eV) at an operating power of 108 W to evaluate the surface chemistry of the samples. The XPS spectra were also calibrated using carbon correction, in agreement with Refs. [25, 26].

# 3. Results

## 3.1. Structural analysis of the electrodeposited overlay coatings

The top-view SE microstructures and EDS spectra of the as-electrodeposited coatings are revealed in Figure 2. The coatings exhibit complete coverage and relatively uniform morphological features, while the EDS profiles indicate only peaks corresponding to Pt and Ru. The EDS-extracted chemical compositions of the nominal Pt, 70Pt-30Ru, 30Pt-70Ru, 10Pt-90Ru, and Ru coatings were 99.4Pt, 71.3Pt-27.6Ru, 31.6Pt-67.8Ru, 11.4-Pt 88.1Ru, and 99.2Ru (in wt%), respectively.

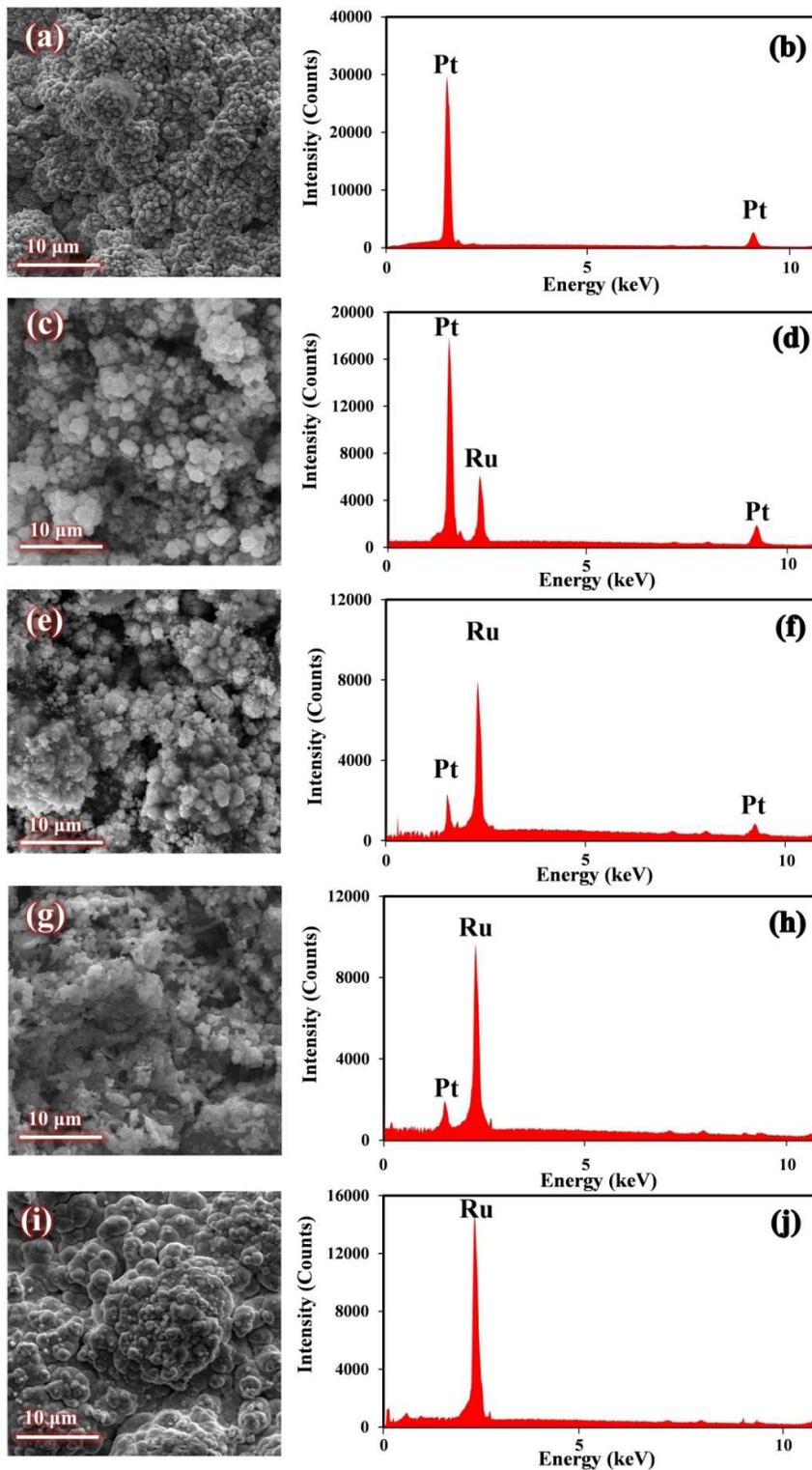

**Figure 2**. SE surface images and EDS profiles of the as-electrodeposited Pt (a, b), 70Pt-30Ru (c, d), 30Pt-70Ru (e, f), 10Pt-90Ru (g, h), and Ru (i, j) coatings.

Figure 3 represents the cross-sectional BSE micrographs of the electrodeposited, annealed samples. After annealing at 1080 ºC for 6 h, the Pt, Ru, and Pt-Ru layers exhibit two distinct contrasts assigned to γ and β phases, serving as the matrix and discontinuous dispersions, respectively. As observed, the proportion of the β phase decreases slightly with increasing the Ru content in the overlays. Table 3 tabulates the EDS-determined compositions (mean ± standard deviation) of the areas marked in Figure 3. The data indicate that the level of Ni in the Pt layer is higher compared to all the Ru-containing layers, but with no significant variations observed among the Ru-containing overlays. The levels of Co and Cr exhibit no consistent dependency on the composition of the electrodeposited layers. However, an increase in the Ru content corresponds to a progressive reduction in the Al content of the overlays, while the concentrations of Y, Ta, and Ti show a gradual increase.

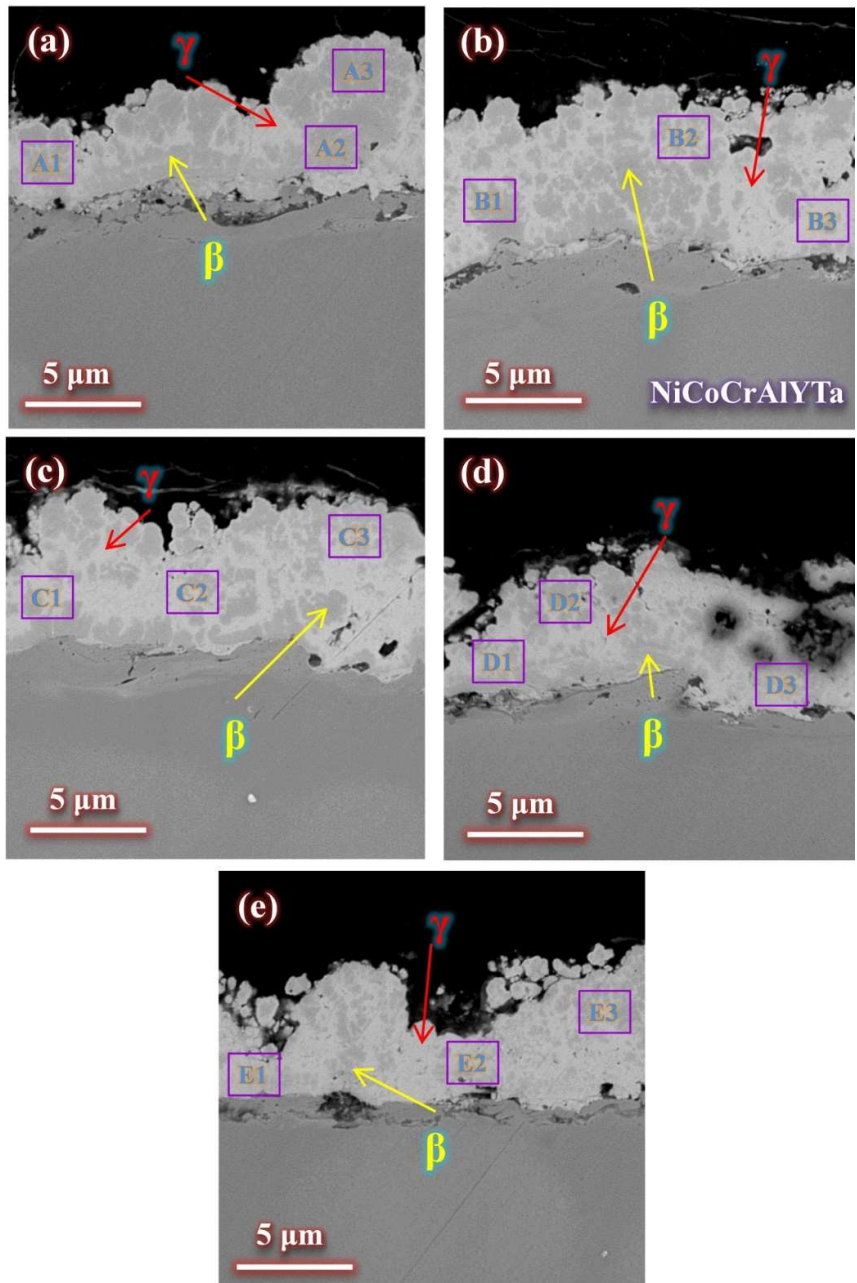

**Figure 3**. Cross-sectional BSE-SEM images for the Pt (a), 70Pt-30Ru (b), 30Pt-70Ru (c), 10Pt-90Ru (d), and Ru (e) deposited samples after annealing at 1080 ˚C for 6 h.

**Table 3.** EDS-extracted compositions of the areas indicated in Figure 3, taken from three different regions for each sample

| Location | Content (wt%) |
| --- | --- |

|   | Pt | Ru | Ni | Co | Cr | Al | Y | Ta | Ti |
|---|---|---|---|---|---|---|---|---|---|
| **A** | 30.2±2.5 | 0.0 | 25.6±1.5 | 11.1±0.5 | 8.4±0.6 | 23.4±2.0 | 0.0 | 0.2±0.1 | 0.3±0.1 |
| **B** | 21.3±1.5 | 11.4±1.5 | 21.4±2.1 | 13.4±1.1 | 11.2±0.5 | 19.6±1.5 | 0.0 | 0.8±0.5 | 0.3±0.1 |
| **C** | 11.5±0.5 | 24.1±1.5 | 20.1±1.8 | 12.2±1.1 | 11.5±0.8 | 18.2±1.5 | 0.2±0.1 | 1.8±0.1 | 0.3±0.1 |
| **D** | 5.2±0.1 | 31.3±2.5 | 23.2±2.1 | 11.3±1.1 | 9.4±0.6 | 16.2±1.2 | 0.3±0.1 | 2.5±0.1 | 0.4±0.1 |
| **E** | 0.0 | 37.4±2.2 | 21.5±1.5 | 12.2±1.1 | 10.4±0.9 | 14.2±1.2 | 0.5±0.1 | 3.2±0.5 | 0.5±0.1 |

### 3.2. Oxidation behavior of the samples

Figure 4 indicates the oxidation kinetic curves of the specimens at 1100 °C. The results depict that the oxidation resistance follows the order: Pt-coated > 70Pt-30Ru-coated > 30Pt-70Ru-coated > 10Pt-90Ru-coated > Ru-coated >> bare NiCoCrAlYTa. In other words, while the electrodeposited coatings significantly enhance the oxidation resistance of the NiCoCrAlYTa-coated substrate, the protection slightly diminishes by increasing the Ru content.

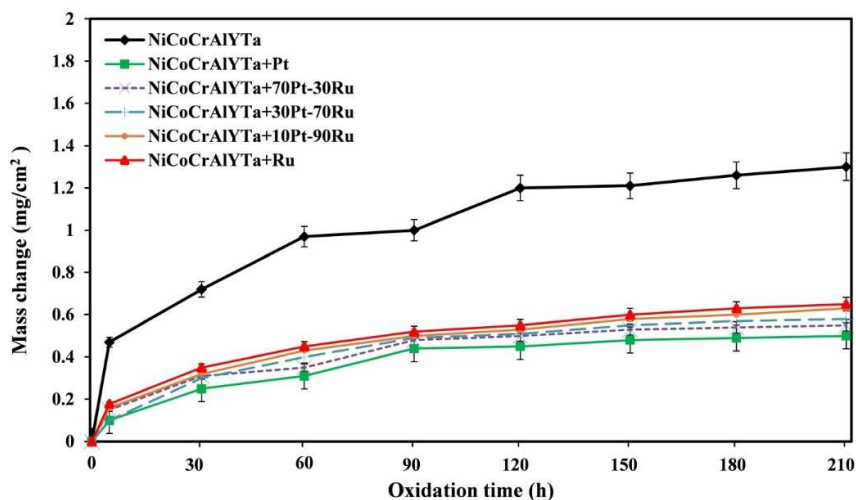

**Figure 4**. Oxidation weight change profile of the samples.

The XRD analysis was conducted on the surfaces after 50 and 200 h of isothermal oxidation (Figure 5). The NiCoCrAlYTa overlay sample exhibits $\alpha$-Al$_2$O$_3$, NiAl$_2$O$_4$, $\gamma/\gamma'$, $\alpha$-Cr, and Cr$_2$O$_3$ phases after 50 h of oxidation. By progression of oxidation to 200 h, diffraction signals for the α-Cr phase disappear, while signals for the other phases increase. For the Pt-Ru modified samples, the α-Cr phase is absent and the oxides formed include α-$Al_2O_3$, along with small amounts of spinel $NiAl_2O_4$. Extending the oxidation duration to 200 h results in a rise in both the intensity and number of the peaks. Over both test periods, the intensity and number of the $NiAl_2O_4$ peaks show an increase, while those of the α-$Al_2O_3$ peaks decrease by increasing the Ru content of the electrodeposited coatings. Compared to the modified samples, the NiCoCrAlYTa overlay sample exhibits weaker α-$Al_2O_3$ signals but stronger spinel phase ones.

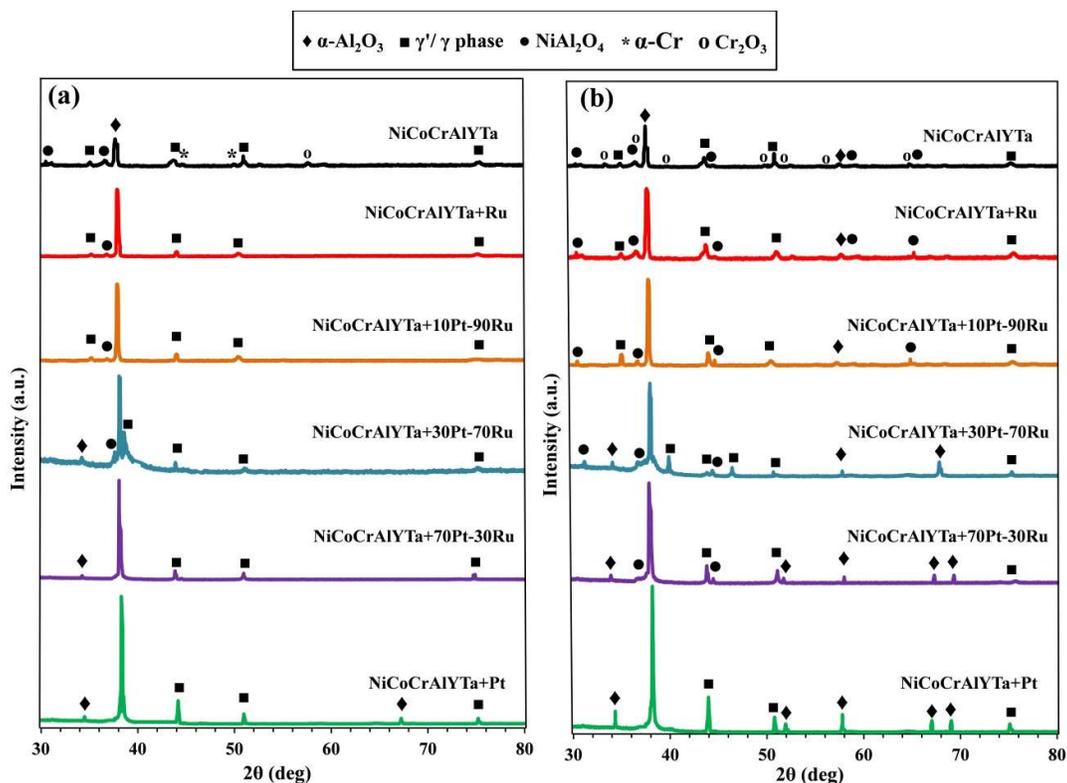

**Figure 5**. XRD profiles of the sample surfaces after oxidation exposure at 1100 °C for 50 (a) and 200 (b) h.

Figure 6 presents the top-view BSE-SEM observations of the Pt and Ru-modified samples after 200 h of oxidation, with the corresponding EDS analysis results summarized in Tables 4 and 5 for distinct locations identified in the micrographs. In the Pt-coated sample, the major phase is characterized to be rich in Al and O, corresponding to $Al_2O_3$, with traces of Ni,

Co, Cr, Y, Ta, and Ti. The microstructure also includes islands enriched in Pt and containing notable amounts of Ni and Al but entirely lacking Ta and Ti. Also, spallation zones and mixed oxides depleted of Ta and Ti are present but in minor quantities. In contrast, the Ru-coated sample exhibits a higher proportion of spallation zones, cracks, and mixed oxides, notably with the mixed oxides containing Ta and Ti.

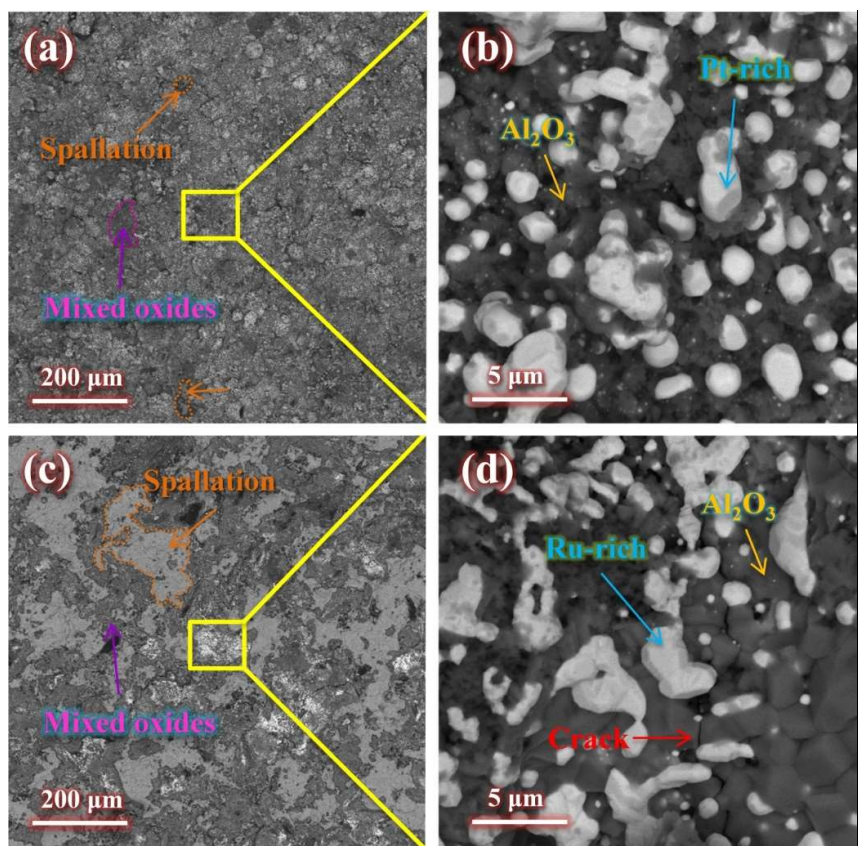

**Figure 6**. Surface BSE-SEM micrographs of the Pt- (a, b) and Ru (c, d)-modified samples after isothermal oxidation exposure at 1100 ºC for 200 h.

**Table 4.** EDS composition of the regions marked in Figure 6 (a, b), based on three repetitions

| Location | Content (wt%) | | | | | | | | | |
|---|---|---|---|---|---|---|---|---|---|---|
| | Pt | Ru | Ni | Co | Cr | Al | O | Y | Ta | Ti |

| | Pt | Ru | Ni | Co | Cr | Al | O | Y | Ta | Ti |
|---|---|---|---|---|---|---|---|---|---|---|
| **Mixed Oxides** | 6.6±0.5 | 0.0 | 14.3±1.1 | 4.3±0.5 | 9.6±0.6 | 25.8±2.2 | 38.8±3.1 | 0.6±0.1 | 0.0 | 0.0 |
| **Al₂O₃** | 0.0 | 0.0 | 1.2±0.1 | 1.4±0.1 | 2.6±0.1 | 33.8±3.1 | 54.8±5.1 | 0.2±0.1 | 4.7±0.1 | 1.3±0.1 |
| **Pt-Rich** | 45.8±3.5 | 0.0 | 27.3±2.5 | 4.1±0.1 | 5.3±0.2 | 13.2±1.1 | 3.6±0.1 | 0.7±0.1 | 0.0 | 0.0 |
| **Spallation** | 0.0 | 0.0 | 43.3±3.5 | 21.1±1.5 | 19.6±1.2 | 9.2±0.1 | 2.7±0.1 | 0.6±0.1 | 3.5±0.1 | 0.0 |

**Table 5.** EDS composition of the regions marked in Figure 6 (c, d), based on three repetitions

| Location | Content (wt%) | | | | | | | | | |
|---|---|---|---|---|---|---|---|---|---|---|
| | Pt | Ru | Ni | Co | Cr | Al | O | Y | Ta | Ti |
| **Mixed Oxides** | 0.0 | 9.8±0.1 | 26.1±2.1 | 4.2±0.1 | 4.1±0.1 | 19.2±1.5 | 29.5±2.5 | 3.4±0.1 | 3.5±0.1 | 0.2±0.1 |
| **Al₂O₃** | 0.0 | 0.2±0.1 | 8.7±0.1 | 1.9±0.1 | 3.1±0.1 | 33.5±3.1 | 52.1±2.5 | 0.5±0.1 | 0.0 | 0.0 |
| **Ru-Rich** | 0.0 | 45.5±2.3 | 28.4±1.5 | 2.7±0.1 | 3.6±0.1 | 5.6±0.1 | 6.1±0.1 | 2.4±0.1 | 5.7±0.1 | 0.0 |
| **Spallation** | 0.0 | 0.0 | 41.4±2.6 | 22.6±1.3 | 21.3±1.1 | 8.4±0.1 | 1.6±0.1 | 0.5±0.1 | 4.2±0.1 | 0.0 |

The cross-sectional BSE-SEM micrographs and mapping of the Pt- and Ru-coated samples after isothermal oxidation at 1100 ºC for 50 h are depicted in Figure 7, revealing four distinct contrasts. The bottom layer (A) corresponds to the NiCoCrAlYTa bond coat layer, which has undergone elemental interdiffusion with the upper layers. Layer B is predominantly composed of Al and O, while this layer in the Ru-coated sample is thicker and contains small dispersions of a secondary phase with a different contrast. In both samples, Layer C mainly consists of Ni, Co, Cr, Al, and O, while this layer in the Ru-coated sample contains Ti in contrast to the Pt-coated sample. The upper islands marked as D are rich in Pt or Ru and include traces of Ni, Co, and Cr. Notably, in the Pt-coated sample, the islands are mainly free of Ta, Ti, and Y but contains some Al, whereas in the Ru-coated sample, the islands are primarily free of Al and Ti but include a minor amount of Ta and Y.

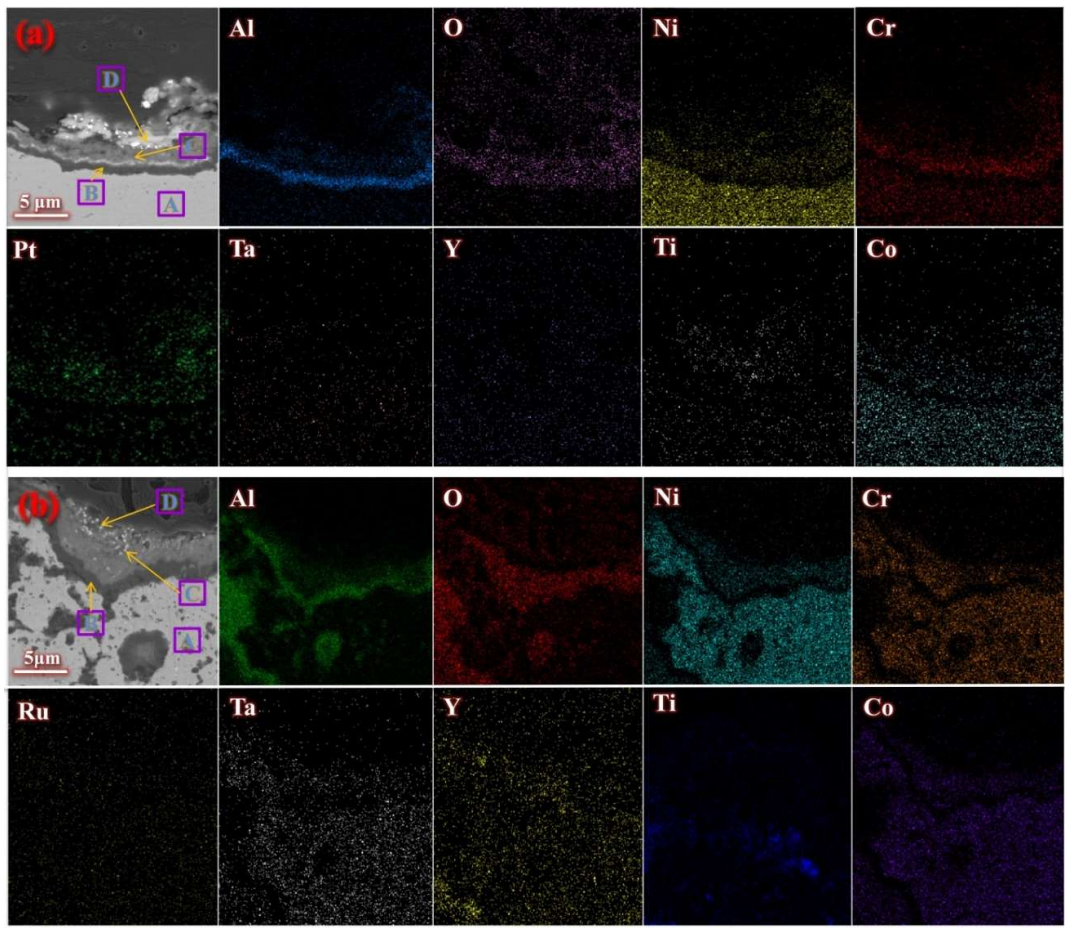

**Figure 7**. Cross-sectional BSE-SEM micrographs and maps for the Pt (a) and Ru (b) coated samples after oxidation at 1100 °C for 50 h.

Figure 8 demonstrates the effect of the Ru content in the electrodeposited overlays on the structural characteristics of the layers formed during oxidation. As observed, the increase of the Ru content leads to a progressive thickening and increased inhomogeneity of the $Al_2O_3$ layer. In the Pt-coated sample, this layer contains small islands attributed to (Ta, Ti)C [7, 27]. However, by increasing the Ru content, the $Al_2O_3$ layer gradually incorporates an additional phase with a contrast distinct from the carbide observed in the Pt-coated sample, assigned to other oxides, along with the development of microcracks. The dispersion of Pt-, Ru-, or Pt-Ru-rich islands within the mixed oxide matrix is also evident in these microstructures.

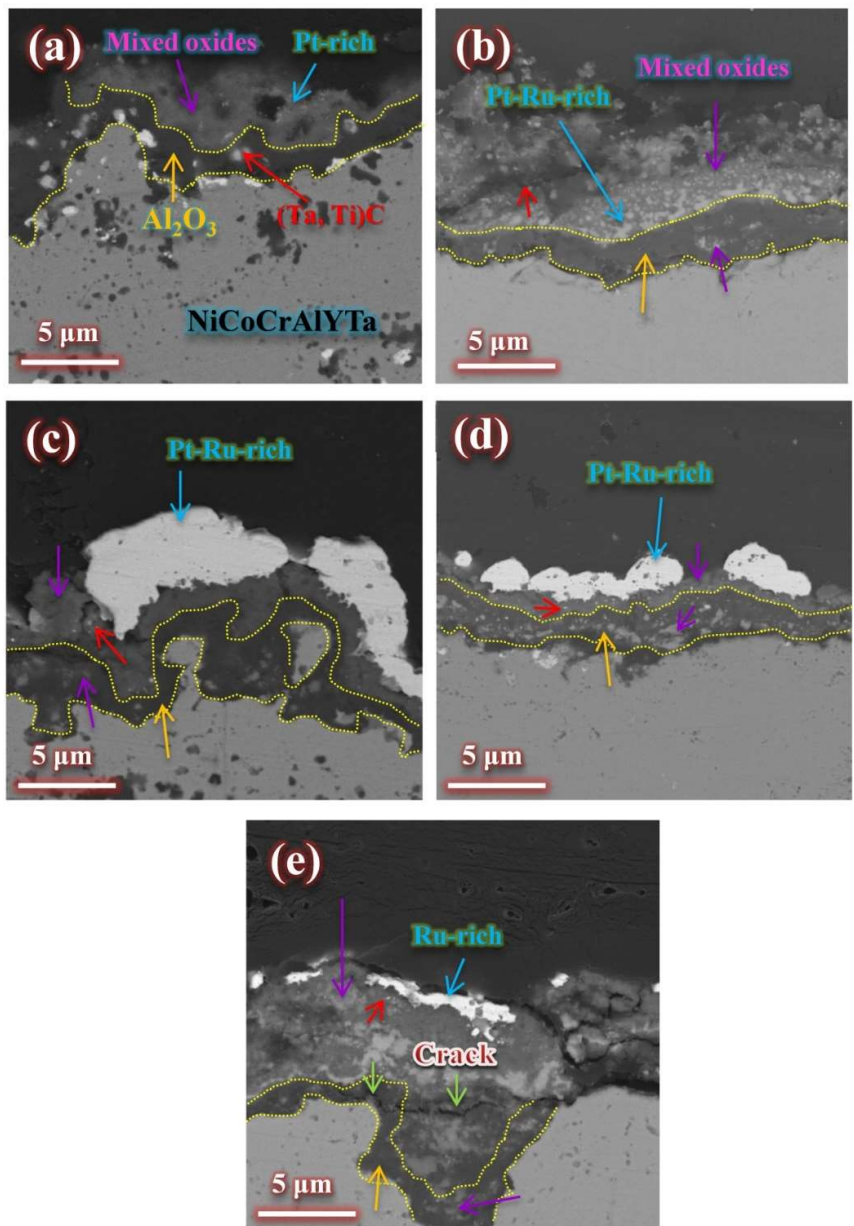

**Figure 8**. Cross-sectional BSE-SEM images of the Pt (a), 70Pt-30Ru (b), 30Pt-70Ru70 (c), 10Pt-90Ru90 (d), and Ru (e) coatings after oxidation at 1100 °C for 50 h.

The XPS profiles of the sample surfaces after oxidation at 1100 ºC for 200 h are revealed in Figure 9. For all the samples, the deconvolution of the high-resolution C 1s scan suggests the dominance of the C1 peak located at 284.9 eV, which can be assigned to C-C, C=C, and C-H bonds [28, 29]. Lower-intensity, higher-energy deconvoluted peaks, C2 and C3, correspond to C-O and C=O bonds, located at 286.3 and 288.3 eV, respectively [30, 31]. These detected bonds are representative of CO and $CO_2$ gases adsorbed during oxidation exposure [32]. The high-resolution scan of Al 2p for all the samples exhibits a broad band ranging from approximately 72 to 76 eV, structured by two primary peaks: Al1 at 72.6 eV and Al2 at 74.3 eV. The Al1 peak corresponds to the dominant Al $2p3/2$ component [33, 34], attributed to metallic Al within the $\gamma'$ phase. The Al2 peak primarily represents the $Al^{3+}$ $2p3/2$ state and is associated with $Al^{3+}$ in $Al_2O_3/NiAl_2O_4$ [35, 36].The high-resolution fitting of the Ni 2p band indicates three peaks at 852.6 (Ni1), 855.1 (Ni2), and 862.2 (Ni3) eV, corresponding to $Ni^0$ $2p_{3/2}$, $Ni^{2+}$ $2p_{3/2}$ and Ni satellite peaks, respectively, compatible with Refs. [37-41]. The peaks of $Ni^0$ and $Ni^{2+}$ are assigned to the $\gamma'$ and $NiAl_2O_4$ spinel phases, respectively. While Cr, Co, Y, and Ti were not detected on the surfaces, Ta was detected only in the Ru-coated sample. Specifically, the $Ta^0$ (Ta1), $Ta^{5+}$ $4f_{7/2}$ (Ta2) and $Ta^{4+}$ $4f_{5/2}$ (Ta3) peaks, centered at 21.9, 26.7, and 28.6 eV, correspond to metallic Ta, $Ta_2O_5$ and $TaO_2$, respectively [42-44] with the dominance of $Ta_2O_5$, realized from the relative intensity of the deconvoluted peaks. The O 1s envelope reflects the dominance of $Al_2O_3$ species, corresponding to $O^{2-}$ at 530.7eV, followed by $NiAl_2O_4$ at 530.9 eV [40, 45, 46] in all the coatings. The O 1s peaks at 531.5 and 529.6 eV confirm the existence of $Ta_2O_5$ and $TaO_2$, respectively, aligning with Ref. [43, 47], in the $NiCoCrAlYTa+Ru$ coating. Also, the peak of 531.5 eV is attributed to adsorbed oxygen species, including $OH^-$ in water and CO in $CO_3^{2-}$[48].

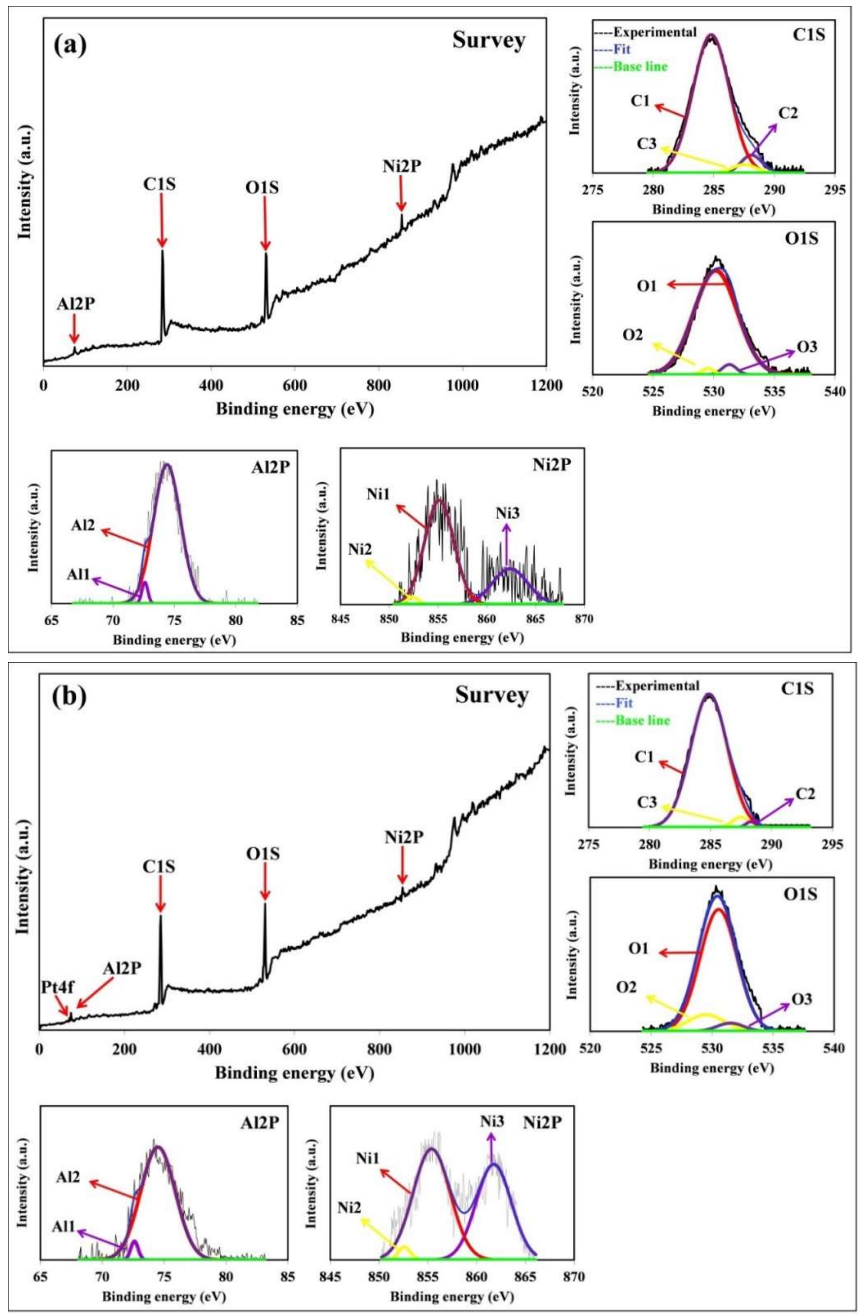

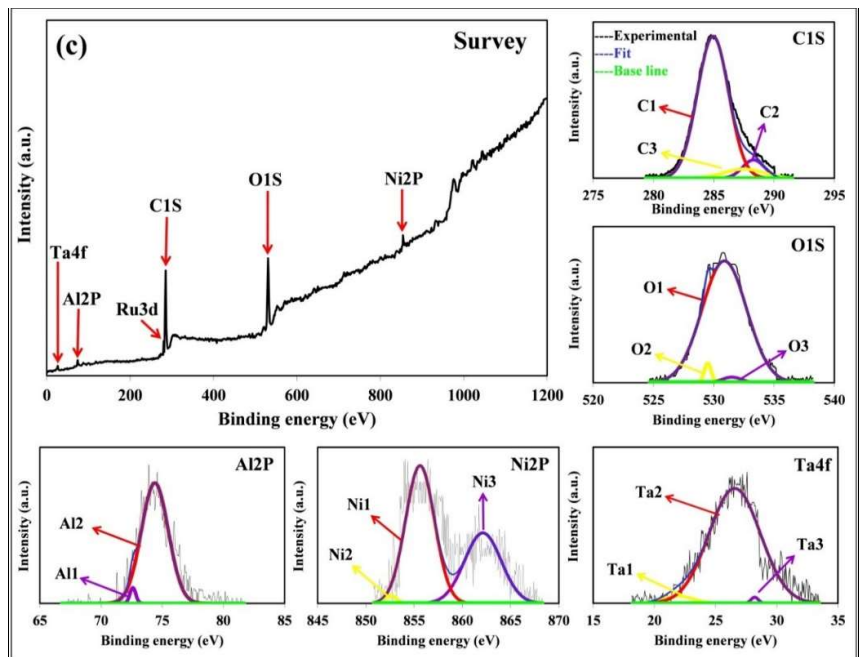

**Figure 9**. Survey and high-resolution XPS spectra for the NiCoCrAlYTa (a), NiCoCrAlYTa+Pt (b), and NiCoCrAlYTa+Ru (c) samples after oxidation at 1100 ºC for 200 h.

## 4. Discussion

The exclusive observation of EDS peaks for Pt and Ru from the as-electrodeposited surfaces (Figure 2), with no detectable impurities, confirms the successful electrodeposition of these elements. Additionally, the uniform surface morphologies observed in Figure 2 further validate the controlled deposition process, achieved through the use of optimal bath chemistry and deposition parameters. According to our recent study [23], electrodeposited Pt, 66Pt-34Ru, and Ru overlays exhibited globular, dendritic, and cauliflower-like surface morphologies, respectively. These morphologies result from the interplay between charge-transfer and diffusion-controlled deposition mechanisms, which align with the findings in the present work from the uniformity viewpoint.

Vacuum-annealing of the electrodeposited samples facilitated the development of the $\gamma$ and $\beta$ phases within the overlays (Figure 3), which is consistent with findings reported in Refs. [5, 49, 50]. These phase structures are driven by the considerable interdiffusion of the elements from the HVOF-deposited NiCoCrAlYTa bond coat into the electrodeposited layers, as corroborated by the data in Table 3. Typically, the $\gamma$ phase in the annealed Pt, Pt-Ru, and Ru overlays is a disordered FCC substitutional solid solution, with compositions based on Ni-Pt, Ni-Pt-Ru, and Ni-Ru, respectively. The $\beta$ phase, richer in Al, is characterized by ordered B2-structured NiPtAl, NiPtRuAl, and NiRuAl intermetallic compounds, acting as a reservoir for Al for forming a protective $Al_2O_3$ layer at elevated temperatures. The other elements listed in Table 3 are dissolved within these two phases [51-53] and/or form separate phases [5, 54], which are albeit difficult to distinguish due to their low quantities.

The variation observed in the compositions and relative phase levels observed in the annealed overlays (Table 3) can be explained by considering the sign and magnitude of the enthalpy of mixing between elemental pairs, which serves as a measure of their affinity [23, 55, 56]. The higher level of Ni diffusion into the pure Pt layer compared to the Ru-containing layers is attributed to the larger negative enthalpy of mixing between Pt and Ni (-5 kJ/mol) compared to Ru and Ni (0). However, the magnitude and difference in their enthalpy of mixing are not large enough to cause meaningful variations in the Ni concentration across the Ru-containing overlays. The irregular and insignificant dependency of the Co and Cr levels on the chemical composition of the electrodeposited layers can be also attributed to their low concentrations in the underlay and their relatively low affinity for the primary elements in the overlays. In contrast, Al exhibits a moderately exothermic enthalpy of mixing with both Pt and Ru, which facilitates its interdiffusion into the overlays. However, the lower concentrations of Al observed with increasing the Ru content are due to the stronger affinity of Pt for Al compared to Ru, with enthalpies of mixing of -44 and -21 kJ/mol for Pt-Al and Ru-Al pairs, respectively. This results in a reduction of the β/γ ratio with an enhancement in the Ru content in the overlays, as Al—a primary constituent of the β phase—becomes less available, as seen in Figure 3. The accumulation trends of Y, Ta, and Ti can be attributed to their highly negative enthalpy of mixing with Pt, which inhibits their interdiffusion, and their moderately negative enthalpy of mixing with Ru, which facilitates their integration into the overlays.

A reciprocal connection of Figures 5-9 and Tables 4 and 5 highlights the microstructures formed during oxidation, thereby suggesting the associated oxidation mechanisms. During the oxidation process, the β phase shown in Figure 3 provides Al for the formation of $Al_2O_3$ and other Al-containing oxides, detected by the XRD and XPS analyses in Figures 5 and 9, respectively. This phase is subsequently depleted of Al, and transforms into an ordered $L1_2$-structured substitutional solid solution (γ′), with primary compositions of $(Ni, Pt)_3Al$, $(Ni, Pt, Ru)_3Al$, and $(Ni, Ru)_3Al$ in the Pt, Pt-Ru, and Ru modified specimens, respectively, in agreement with Refs. [53, 57]. By progression of oxidation, γ′ becomes depleted of Al and transforms into γ. As well as Al, other elements, including Ni, Co, and Cr, undergo oxidation, forming spinel phases, particularly $NiAl_2O_4$ or $(Ni, Co, Cr)Al_2O_4$, and/or other oxides like $NiCr_2O_4$, and $CoAl_2O_4$ [50, 53]. Based on the XPS analysis (Figure 9), $NiAl_2O_4$ is present in the surface layers, while the XRD (Figure 5) and EDS (Figures 6 and 7, and Tables 4 and 5) analyses showed the presence of mixed oxides in the subsurface layers. Accordingly, γ/γ′ becomes enriched with Pt and Ru, as Pt and Ru are resistance to oxidation at 1100 °C [58], forming metallic islands rich in these noble elements, as indicated in Figures 6-9 and Tables 4 and 5, in agreement with Ref. [50, 59]. The appearance of $NiAl_2O_4$ and/or $(Ni, Co, Cr)Al_2O_4$ in the XRD analysis (Figure 5) suggests the dominance of the spinel phase in the mixed oxide zones, assuming other oxides are also present. If distributed within a phase, Ta can trap Ti and C diffusing from the superalloy toward the oxide scale through the formation of (Ta, Ti)C carbides [60, 61]; otherwise, Ti oxidation imposes detrimental effects on the overall oxidation resistance [54, 62].

A thorough comparison of the microstructures obtained in the various samples after oxidation (Figures 5-9 and Tables 3 and 4) can explain the observed oxidation mass gain and resistance (Figure 4), as schematically summarized in Figure 10. The most significant factor contributing to the enhanced oxidation resistance with increasing the Pt content is the evolution of the $\alpha$-$Al_2O_3$ layer, which serves the primary component of oxidation protection in this system [7, 63]. As demonstrated in the XRD analysis (Figure 5), the relative amount of the $\alpha$-$Al_2O_3$ phase is reduced by increasing the Ru content in the specimens, while the microscopic evaluations (Figures 6-8) reveal a deterioration in the compactness and homogeneity the $Al_2O_3$ layer. This aligns well with the observed enhancement in Al accommodation and the $\beta$ proportion with increasing the Pt content (Figure 3 and Table 3). Notably, the $Al_2O_3$ layers in the samples coated with the Ru-containing overlays incorporate mixed oxides, which are well-documented to have a detrimental effect on oxidation [53, 64, 65].

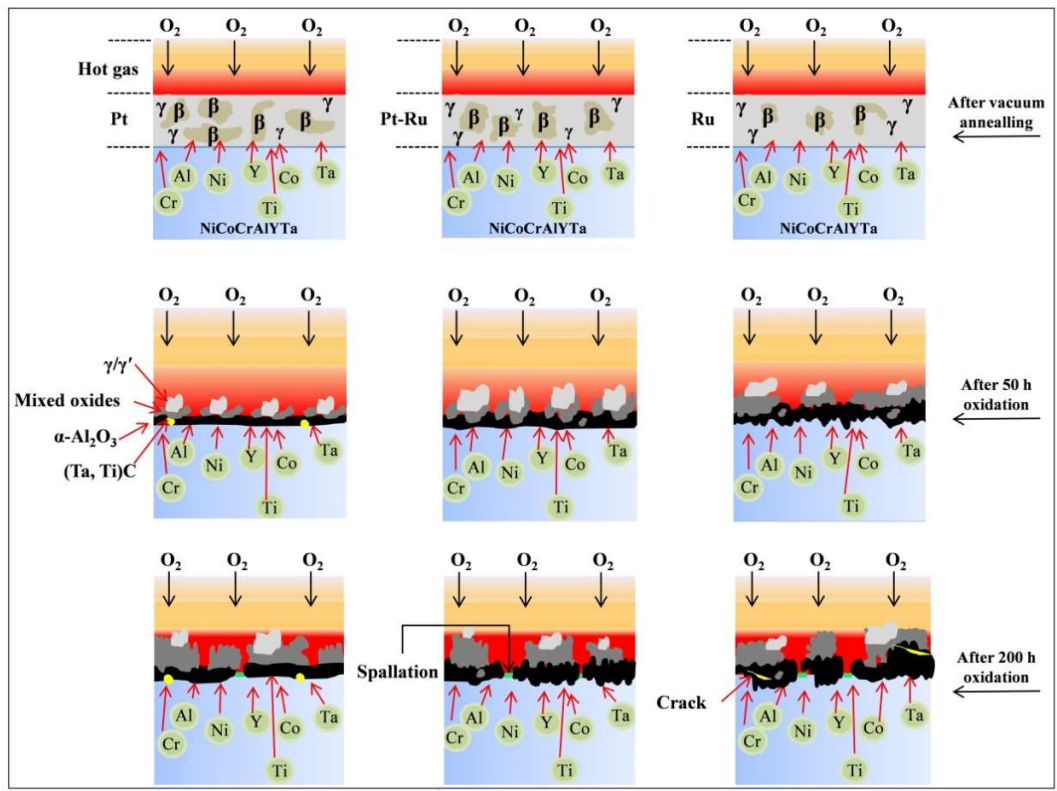

Figure 10. Mechanisms representing the oxidation evolution in the samples.

The role of Y, Ta, and Ti accommodation in the achieved microstructures further supports the enhancement observed in oxidation protection with increasing the Pt content. Y enhances adhesion and reduces the growth rate of the TGO layer [3, 66]. However, the enhanced accumulation of Y in the metallic islands with increasing the Ru content (Figures 6 and 7 and Table 4 and 5) negates its positive effect. Conversely, the reduced tendency of Y to diffuse into the islands with increasing the Pt content directs it toward the oxide layer, particularly the $Al_2O_3$ layer, leading to its beneficial contribution. Additionally, as indicated in Tables 3 and 4 as well as Figures 7 and 9, with the increased Ru amount, Ta atoms accommodate within the metallic islands and mixed oxide zones, while Ti, which has a high oxidation tendency, tends to accommodate in the mixed oxide zones. In contrast, in the Pt-coated sample, both Ti and Ta are expelled from the metallic islands, enabling the formation of (Ta, Ti)C in the oxide zones and minimizing the negative effect of Ti on oxidation protection. Additionally, the incorporation of Ta in the TGO layer of the Pt-coated sample enhances Al diffusivity [41, 67] while decreasing both the inward diffusion of oxygen [68] and the activity of $O^{2-}$ [69], slowing the formation of $Al_2O_3$. As a result, the suppressed excessive growth of the TGO layer improves its adhesion and enhances oxidation resistance. The partitioning of these three elements (Ta, Y, and Ti) in the noted phases aligns well with the EDS results of the annealed samples (Table 3).

The existence of oxides other than alumina within the TGO layer of the Ru-coated samples is responsible for crack formation and spallation in the high Ru samples (Figures 6d and 8e), as can be elucidated via the Pilling–Bedworth ratio (PBR) and coefficient of thermal expansion (CTE) mismatch. PBR compares the volume of the oxide that forms to the volume of the metal consumed to form it. A PBR of 1.28 for $Al_2O_3$ [70], as the dominant phase of the high Pt samples, is consistent with a protective oxide layer and desirable adhesion to the substrate. By increasing the Ru content of the overlays, and accordingly promoting the evolution of other oxides, primarily $NiAl_2O_4$ (PBR $\approx$ 2.62 [71]), the oxide's volume significantly exceeds that of the original metal. As a result, during oxide formation, the oxide attempts to expand much more than the underlying metal can accommodate, while the metal substrate restricts this expansion. This mismatch generates substantial compressive stresses within the oxide layer, rendering it unstable and leading to spallation and cracking. On the other hand, the CTE values for NiCoCrAlYTa, $Al_2O_3$, and $NiAl_2O_4$ are $18.4 \times 10^{-4}$ [72], $8.5 \times 10^{-6}$ [73], and $5.1 \times 10^{-6}$ [74] $K^{-1}$, respectively. The increase in CTE mismatch due to the formation of $NiAl_2O_4$ in comparison to alumina further supports the observed cracking and spallation in the high Ru samples, as thermal cycling induces stress build-up, and upon heating and cooling, mismatch strains cause cracking, delamination, or spallation of the oxide layer.

## 5. Conclusions

This mechanistic study concluded that electrodeposited overlays within the Pt-Ru alloy system significantly enhance the high-temperature oxidation resistance of NiCoCrAlYTa bond coats in TBC systems. However, increasing the Ru content slightly reduces oxidation resistance due to the interplay of kinetic and thermodynamic interactions among the overlay, bond coat, and substrate elements. Nevertheless, given Ru's significantly lower cost compared to Pt, a comprehensive economic analysis, alongside durability evaluation, is essential to explore Ru as a lower-cost alternative or supplementary option to traditional Pt-modified coatings while maintaining or enhancing performance.